\documentclass[fleqn,twoside]{article}
\usepackage{espcrc2}
\usepackage{graphicx}

\newcommand{\nature}[3]{Nature {\bf B#1}, #3 (#2)}

\newcommand{\apj}[3]{Astrophys.\ J.\ {\bf #1}, #3 (#2)}
\newcommand{\prl}[3]{Phys.\ Rev.\ Lett. {\bf #1}, #3 (#2)}
\newcommand{\prd}[3]{Phys.\ Rev.\ {\bf D#1}, #3 (#2)}

\title{Generalized Cardassian Expansion: a Model in which the Universe is Flat,
Matter Dominated, and Accelerating\\
}

\author{Katherine Freese\address[MCTP]{Michigan Center for Theoretical
Physics, \\ Physics Department, \\ University of Michigan, \\ Ann Arbor,
MI 48109, USA \\
}}

\begin{document}

\begin{abstract}

The Cardassian universe is a proposed modification to the Friedmann
Robertson Walker equation (FRW) in which the universe is flat, matter
dominated, and accelerating.  In this presentation, we generalize the
original Cardassian proposal to include additional variants on the FRW
equation; specific examples are presented.

In the ordinary FRW equation, the right hand side is a linear function
of the energy density, $H^2 \sim \rho$.  Here, instead, the right hand
side of the FRW equation is a different function of the energy density,
$H^2 \sim g(\rho)$.  This function returns to ordinary FRW at early
times, but modifies the expansion at a late epoch of the universe.  The
only ingredients in this universe are matter and radiation: in
particular, there is NO vacuum contribution.  Currently the modification
of the FRW equation is such that the universe accelerates; we call this
period of acceleration the Cardassian era.  The universe can be flat and
yet consist of only matter and radiation, and still be compatible with
observations.  The energy density required to close the universe is much
smaller than in a standard cosmology, so that matter can be sufficient
to provide a flat geometry.  The new term required may arise, e.g., as a
consequence of our observable universe living as a 3-dimensional brane
in a higher dimensional universe.  The Cardassian model survives several
observational tests, including the cosmic background radiation, the age
of the universe, the cluster baryon fraction, and structure formation.
As will be shown in future work, he predictions for observational tests
of the generalized Cardassian models can be very different from generic
quintessence models, whether the equation of state is constant or time
dependent.
\vspace{1pc}
\end{abstract}

\maketitle

Recent observations of Type IA Supernovae \cite{SN1,SN2} as well as
concordance with other observations (including the microwave background
and galaxy power spectra) indicate that the universe is accelerating.
Many authors have explored a cosmological constant, a decaying vacuum
energy \cite{fafm,frieman}, and quintessence \cite{stein,caldwell,huey}
as possible explanations for such an acceleration.

Recently we proposed Cardassian expansion \cite{freeselewis} (hereafter
Paper I)
\footnote{The name Cardassian refers to a humanoid race in Star Trek
whose goal is accelerated expansion of their evil empire.
This race looks foreign to us and yet is made entirely of matter.}
 as an explanation for acceleration which invokes no vacuum
energy whatsoever.  In our model the universe is flat and accelerating,
and yet consists only of matter and radiation.  Previously we considered
the addition of a new term to the right hand side of the FRW equation:
\begin{equation}
\label{eq:new}
H^2 = A \rho + B \rho^n 
\end{equation}
where energy density $\rho$ contains
only matter and radiation (no vacuum) and $n$ is a time
independent number with
\begin{equation}
n<2/3 .
\end{equation}
Here
$H = \dot R / R$ is the Hubble constant (as a function of time) and
$R$ is the scale factor of the universe.  
In the usual FRW equation
$B=0$.  To be consistent with the usual FRW result, we take $A={8\pi
\over 3 m_{pl}^2}$. 
The new term is initially
negligible, and only comes to dominate at redshift $z \sim {\cal O}(1)$.
Once it dominates, it causes the universe to accelerate,
as discussed further below.

\section{Generalized Cardassian Models}

Here we wish to generalize this proposal to other functions on the right
hand side of the FRW equation.  Pure matter (or radiation) alone can
drive an accelerated expansion if the Friedmann Robertson Walker (FRW)
equation is modified to become
\begin{equation}
\label{eq:general}
H^2 = g(\rho) ,
\end{equation}
We take $g(\rho)$ to be
a function of $\rho$ that returns simply to $\rho$ at early 
epochs, but that can drive an accelerated expansion in the recent past
of the universe at $z<{\cal O}(1)$.
We take the usual energy conservation:
\begin{equation}
\label{eq:energy}
\dot \rho + 3H (\rho + p) = 0 ,
\end{equation}
which gives the evolution of matter:
\begin{equation}
\rho_M = \rho_{M,0}(R/R_0)^{-3} .
\end{equation}
Here subscript $0$ refers
to today.  Eqs.(\ref{eq:general}) and (\ref{eq:energy})
contain the complete information of the two Friedmann equations.
 
We note here that the geometry is flat, as required by measurements of
the cosmic background radiation \cite{boom}, so that there are no
curvature terms in the equation.  There is no vacuum term in the
equation.  This paper does not address the cosmological constant
($\Lambda$) problem; we simply set $\Lambda=0$.

The simplest example of this type of behavior is the sum of two terms:
\begin{equation}
H^2 = \rho + f(\rho)
\end{equation}
where $f(\rho)$ is a different function of $\rho$.  

As mentioned above, in Paper I, the specific form of $f(\rho)$ that we
considered was
$H^2 = A \rho + B \rho^n $ 
with $n< 2/3$ and $n$ constant in time.
Another way to write this equation is
\begin{equation}
\label{eq:new2}
H^2 = A \rho [1 + ({\rho \over \rho_{car}})^{n-1}] .
\end{equation}
The first term inside the bracket dominates initially but the
second term takes over once the energy density has dropped to
the value $\rho_{car}$.
Here, $\rho_{car}$ is the energy density at which the
two terms are equal: the
ordinary energy density term on the right hand side of the FRW
equation is equal in magnitude to the new term.
Hence there are two parameters in the model: one can
take them to be $B$ and $n$,
or equivalently, $\rho_{car}$ and $n$, or equivalently,
$z_{car}$ and $n$.

The new term in the equation (the second term on the right hand side) is
initially negligible.  It only comes to dominate recently, at the
redshift $z_{car} \sim O(1)$ indicated by the supernovae observations.
Once the second term dominates, it causes the universe to accelerate.
When the new term is so large
that the ordinary first term can be neglected, we find
\begin{equation}
R \propto t^{2 \over 3n}
\end{equation}
so that the expansion is superluminal (accelerated) for $n<2/3$.
As examples, for $n=2/3$ we have $R \sim t$;
for $n=1/3$ we have $R \sim t^2$; and for $n=1/6$ we have $R \sim t^4$.
The case of $n=2/3$ produces a term in the FRW
equation $H^2 \propto R^{-2}$; such a term looks similar to a curvature
term but is generated here by matter in a universe with a flat geometry.
Note that for $n=1/3$ the acceleration is constant, for $n>1/3$ the
acceleration is diminishing in time, while for $n<1/3$ the acceleration
is increasing (the cosmic jerk).

Note that the parameter $B$ here is chosen to make the second
term kick in at the right time to explain the observations.
As yet we have no explanation of the
coincidence problem; i.e., we have no explanation for the timing of
$z_{car}$.  Such an explanation would arise if we had a reason for the
required mass scale of $B$; such an explanation may arise in the context
of extra dimensions.

We were motivated to study an equation of this form by the 
work of Chung and Freese \cite{cf} who showed that terms
of the form $\rho^n$ can arise as a consequence of embedding
our observable universe as a brane in extra dimensions.

\section{Examples of Alternative FRW Equations}

We wish to mention here some alternative forms of $g(\rho)$
in Eq.(3).
Wang, Freese, Frieman, and Gondolo \cite{wang} are studying 
three Cardassian alternatives:

\noindent
1)  A simple generalization of Eq.(\ref{eq:new}) is:
\begin{equation}
H^2 = A \rho [1+ ({\rho/\rho_{car}})^{q(n-1})]^{1/q} .
\end{equation}
Here, $q>0$.  As before, we require $n<2/3$.
The right hand side  returns to $A \rho$ (the ordinary
FRW equation) at early times, but becomes
$\rho^n$ at late times, just as in Eq.(\ref{eq:new2}).
However, the cross over time period during which the
two terms are roughly comparable is different here.

\noindent
2) Another possibility is
\begin{equation}
\label{eq:poly}
H^2 = D [1 + (\rho/\rho_{car})^q]^{1/q} .
\end{equation}
This example can have a particularly interesting
equation of state.  Gondolo and Freese \cite{gondolo}
are considering treating the right hand side of
Eq.(\ref{eq:poly}) as a single fluid. Then this fluid
behaves as a polytrope of negative index:
\begin{equation}
p \propto - ({\rho \over \rho_{car}})^{1-q},
\end{equation}
which corresponds to a polytrope
$p= K \rho^{1+1/N}$ with negative index $N=-1/q$ and
negative pressure ($K<0$).

\noindent
3) A third possibility modifies the simplest Cardassian
proposal with a logarithm:
\begin{equation}
H^2 = A \rho + B \rho^n {\rm log}^q \rho .
\end{equation}
Many other possibilities for the function $g(\rho)$
in Eq.(3) exist.

As will be shown in future work \cite{wang}, the predictions
for observational tests of these models can be very different
from generic quintessence models whether the equation of
state is constant or time dependent.

\section{The simplest Cardassian Model: FRW with additional
$\rho^n$ term}

For the rest of this presentation, we study specifically
the case where $g(\rho) = A \rho + B \rho^n$ for constant
$n<2/3$.  This is the case that was studied in Paper I.
We use it to illustrate the basic properties of a Cardassian
model.  

\subsection{\it What is the Current Energy Density of the Universe?}

Observations of the cosmic background radiation show that the geometry
of the universe is flat with $\Omega_0=1$.  In the Cardassian model we
need to revisit the question of what value of energy density today,
$\rho_0$, corresponds to a flat geometry.  We will show that the energy
density required to close the universe is much smaller than in a
standard cosmology, so that matter can be sufficient to provide a flat
geometry.

From evaluating Eq.(\ref{eq:new}) today, we have
\begin{equation}
\label{hubbletoday}
H_0^2 = A \rho_0 + B \rho_0^n .
\end{equation}
The energy density $\rho_0$ that satisfies Eq.(\ref{hubbletoday})
is, by definition, the critical density.
We can solve Eq.(\ref{hubbletoday}) to 
find that the critical density $\rho_c$ has been modified from
its usual value, i.e., the number has changed.
We find
\begin{equation}
\rho_c = \rho_{c,old} \times F(n)
\end{equation}
where
\begin{equation}
\label{eq:F}
F(n) = [1+(1+z_{car})^{3(1-n)}]^{-1}
\end{equation}
and
\begin{equation}
\rho_{c,old} = 1.88 \times 10^{-29} h_0^2 {\rm gm/cm^{-3}}
\end{equation}
and $h_0$ is the Hubble constant today in units of 100 km/s/Mpc.


In the (simplest) Cardassian model with new term $\rho^n$,
the value of the critical density can be much lower
than previously estimated.  Since $\Omega_0 =1$ today,
we have today's energy density as $\rho_0 = \rho_c$
as given above\footnote{An alternate possible definition
would be to keep the standard value of $\rho_c$ and discuss
the contribution to it from the two terms on the
right hand side of the modified FRW equation.
Then there would be contribution to $\Omega$ from
the $\rho$ term and another contribution from the $\rho^n$
term with the two terms adding to 1.  This is the approach
taken when one discusses a cosmological constant in lieu
of our second term.  However, the situation here is different
in that we have only matter in the equation.  The disadvantage
of this second choice of definitions would be that the value
of the energy density today equal to $\rho_c$ equal to $\rho_c$
according to this second definition would not correspond
to a flat geometry.}.

For the past ten years, a multitude of observations has pointed towards
a value of the matter density $\rho_o \sim 0.3 \rho_{c,old}$.  The
cluster baryon fraction \cite{white,evrard} as well as the observed
galaxy power spectrum are best fit if the matter density is 0.3 of the
old critical density.  Recent results from the CMB \cite{boom,dasi} also
obtain this value.  In the standard cosmology this result implied that
matter could not provide the entire closure density.  Here, on the other
hand, the value of the critical density can be much lower than
previously estimated.  Hence the cluster motivated value for $\rho_o$ is
now {\it compatible} with a closure density of matter, $\Omega_o =1$,
all in the form of matter.  

For example, if $n= 0.6$ with $z_{car} =1$,
or if $n=0.2$ with $z_{car} = 0.4$, a critical density of matter
corresponds to $\rho_o \sim 0.3 \rho_{c,old}$, as required by the
cluster baryon fraction and other data.  
If we assume that the value
$\rho_o = 0.3 \rho_{c,old}$ is correct, for a given value of $n$ (that
is constant in time) we can compute the value of $z_{car}$ for our model
from Eq.(\ref{eq:F}).  Henceforth we shall use these combinations
of parameters.

\subsection{\it Other observational tests}

As discussed in Paper I, the simplest Cardassian model with an
additional term $\rho^n$ satisfies many observational constraints: the
universe is somewhat older, the first Doppler peak in the microwave
background is slightly shifted, early structure formation ($z>1$) is
unaffected, but structure will stop growing sooner. In addition the
modifications to the Poisson equation will affect cluster abundances and
the ISW affect in the CMB.

\subsection{\it Comparing to Quintessence}

We note that, with regard to observational tests, one can make a
correspondence between the $\rho^n$ Cardassian and Quintessence models
for constant $n$; we stress, however, that the two models are entirely
different. Quintessence requires a dark energy component with a specific
equation of state ($p = w\rho$), whereas the only ingredients in the
Cardassian model are ordinary matter ($p = 0$) and radiation ($p =
1/3$). However, as far as any observation that involves only $R(t)$, or
equivalently $H(z)$, the two models predict the same effects on the
observation.  Regarding such observations, we can make the following
identifications between the Cardassian and quintessence models: $n
\Rightarrow w+1$, $F\Rightarrow \Omega_m$, and $1-F \Rightarrow
\Omega_Q$, where $w$ is the quintessence equation of state parameter,
$\Omega_m= \rho_m/\rho_{c,old}$ is the ratio of matter density to the
(old) critical density in the standard FRW cosmology appropriate to
quintessence, $\Omega_Q= \rho_Q/\rho_{c,old}$ is the ratio of
quintessence energy density to the (old) critical density, and F is
given by Eq.(\ref{eq:F}).  In this way, the Cardassian model with
$\rho^n$ can make contact with quintessence with regard to observational
tests.

\subsection{\it Best Fit of Parameters to Current Data}

We can find the best fit of the Cardassian parameters $n$ and $z_{car}$
to current CMB and Supernova data.  The current best fit is obtained for
$\rho_o = 0.3 \rho_{c,old}$ (as we have discussed above) and $n<0.4$
(equivalently, $w < -0.6$) \cite{bean,hm}.  
In Table I one can see the values of
$z_{car}$ compatible with this bound, as well as the resultant age of the
universe.  
As an example, for $n= 0.2$ (equivalently, $w=-0.8$), 
we find that $z_{car} = 0.42$.
Then the position of the first Doppler peak is shifted by a factor of
1.12.  The age of the universe is 13 Gyr.  The cutoff energy density is
$\rho_{cutoff} = 2.7 \rho_c$, so that the new term is important only for
$\rho < \rho_{cutoff} = 2.7\rho_c$.  Hence, 
the Cardassian term won't affect the physics of the Earth or solar
system in any way.

\section{\it Discussion}

We have presented $H^2 = g(\rho)$ as a modification to the FRW equations
 in order to suggest an explanation of the recent acceleration of the
 universe.  In the Cardassian model, the universe can be flat and yet
 matter dominated.  We have found that the new Cardassian modifications
 can dominate the expansion of the universe after $z_{car} =
 \mathcal{O}$$(1)$ and can drive an acceleration.  We have found that
 matter alone can be responsible for this behavior.  The current value
 of the energy density of the universe is then smaller than in the
 standard model and yet is at the critical value for a flat geometry.
We reported on results for the simplest Cardassian case of Eq.(1): 
Structure formation is unaffected before $z_{car}$.  The age of the
 universe is somewhat longer.  The first Doppler peak of the cosmic
 background radiation is shifted only slightly and remains consistent
 with experimental results.  Such a modified FRW equation may result
 from the existence of extra dimensions.  Further work is required to
 find a simple fundamental theory responsible for Eq.({\ref{eq:new}).
 In this presentation, generalized cardassian models were discussed.  As
 will be shown in future work, the predictions for observational tests
 of these models can be very different from generic quintessence models,
 whether the equation of state is constant or time dependent.

\section{Acknowledgments}

This paper reflects work with collaborators Matt Lewis, Paolo Gondolo,
Yun Wang, and Josh Frieman.  K.F. thanks T. Baltz, D. Chung,
R. Easther, G. Evrard, P. Gondolo, W. Hu, L. Hui, W.
Kinney, R. Wechsler, and especially J. Liu for many useful
conversations and helpful suggestions.  We acknowledge support from the
Department of Energy via the University of Michigan, and thanks the
Kavli Institute for Theoretical Physics at Santa Barbara.


\end{document}